\documentclass[aps,prl,twocolumn,superscriptaddress,preprintnumbers,floatfix,nofootinbib]{revtex4}
\usepackage{graphicx}
\usepackage{epsfig}

\begin{document}

\newcommand{\be}{\begin{eqnarray}}
\newcommand{\ee}{\end{eqnarray}}
\newcommand\del{\partial}
\newcommand{\nn}{\nonumber } 
\newcommand{\tK}{Z_{n=1}}
\newcommand{\re}{\mathrm{Re}}
\newcommand{\im}{\mathrm{Im}}


\title  {Chiral Symmetry Breaking and the Dirac Spectrum at Nonzero Chemical
  Potential} 

\author{J.C. Osborn}
\affiliation{Physics Department, Boston University,
Boston, MA 02215, USA}
\author{K. Splittorff}
\affiliation{Nordita, Blegdamsvej 17, DK-2100, Copenhagen {\O}, Denmark}
\author{J.J.M. Verbaarschot}
\affiliation{Department of Physics and Astronomy, SUNY, Stony Brook,
 New York 11794, USA}

\date   {\today}

\begin  {abstract}

The relation between the spectral density of the QCD Dirac operator at
nonzero baryon chemical potential and the chiral condensate is investigated.
We use the analytical result for the eigenvalue density
in the microscopic regime which shows oscillations with a period 
that scales as $1/V$ and an amplitude that diverges exponentially with the
volume $V=L^4$.
We find that the discontinuity of the chiral condensate is
due to the whole oscillating region rather than to an accumulation of
eigenvalues at the origin. 
These results also extend beyond the microscopic regime to chemical
potentials $\mu \sim 1/L$.

\end {abstract}

\maketitle

\noindent
{\it Introduction.} 
One of the salient features of QCD at low energy is
the spontaneous breaking of chiral symmetry characterized by a 
discontinuity of the chiral condensate. 
More than two decades ago it was realized 
by Banks and Casher \cite{BC} that the discontinuity of the chiral 
condensate at zero quark mass is proportional to the 
eigenvalue density of the QCD Dirac operator.
This relation establishes that the eigenvalue spectrum of the anti-Hermitian
Dirac operator 
becomes dense at the origin of the imaginary axis in the thermodynamic limit.
The Banks-Casher relation is of substantial practical value
for nonperturbative numerical studies of QCD.
It allows one to extract
the chiral condensate directly from the spectral density.

At nonzero baryon chemical potential, $\mu$, 
the Euclidean Dirac operator, $D \equiv D_\eta\gamma_\eta +
\mu\gamma_0$, is non-Hermitian 
so that the support of its spectrum is a 2 dimensional
domain in the complex plane. 
In this case we will show that the discontinuity of the
chiral condensate at zero mass is not due to the accumulation of
eigenvalues near zero.

The QCD partition function at zero temperature
does not depend on the baryon chemical potential,
\be
Z_{N_f}(m;\mu) = Z_{N_f}(m;\mu=0), \qquad {\rm for} \qquad \mu < \mu_c,
\ee
where $\mu_c$ is the mass per unit quark number of 
the lightest excitation with nonzero quark number. 
Here and below
we only consider the case with $N_f$ quark flavors with equal 
mass $m$. 
Therefore, the chiral
condensate given by 
\be\label{Sigma-def}
\Sigma_{N_f}(m) =\frac{1}{N_f V}\partial_m \log Z_{N_f}(m;\mu),
\ee
remains equal to its value at $\mu = 0$  for $\mu <\mu_c$.
Our aim is to understand this behavior from the spectrum of the
Dirac operator.

We will consider two types of gauge field averages,
quenched averages where the determinant
of the Dirac operator is not included in the average and
unquenched averages which include the fermion determinant. 
The quenched spectral density can be expressed as
\be 
\rho_Q(x,y;\mu) = \langle \sum_k \delta^2(x+iy - z_k) \rangle,
\ee
where the eigenvalues of the Dirac operator are given by $z_k$,
and the brackets denote the average over the Yang-Mills action.
The eigenvalue density of full QCD includes the fermion determinant in
the average  
\be \label{rhoNf-def}
 \rho_{N_f}(x,y,m;\mu) =
\frac{
 \langle \sum_k \delta^2(x+iy - z_k)\,
 {\det}^{N_f}( D+m ) \rangle
}{\langle {\det}^{N_f}( D+m ) \rangle}
.
\ee
Strictly speaking, since $\rho_{N_f}$ is in general complex, this is 
not a density.
Due to chiral symmetry the eigenvalues
occur in pairs $\pm z_k$ so that 
$ \rho_{N_f}(x,y,m;\mu)= \rho_{N_f}(-x,-y,m;\mu)$. 
A second reflection symmetry which holds only after averaging over
the gauge field configurations is that
$\rho_{N_f}^*(x,y,m;\mu)= \rho_{N_f}(x,-y,m;\mu)$.
Because the fermion determinant vanishes for $z_k = \pm m$
we expect that $\rho_{N_f}(x=\pm m,y=0,m;\mu) = 0$. 
The chiral condensate in the chiral limit can be expressed as
\be
\Sigma 
= \lim_{m\to 0} \lim_{V\to\infty}
\frac 1{V} \int dx\,dy \, \frac{\rho_{N_f}(x,y,m;\mu)}{x+iy +m}.
\label{psibarpsi}
\ee
At zero chemical potential the above quantity is known to be proportional to
the density of the imaginary eigenvalues at zero \cite{BC}.
This happens because then  $\rho_{N_f}(x,y,m;\mu=0) \propto \delta(x)$. 
When $\mu\ne0$ the eigenvalues spread into the complex plane and this argument
no longer holds.
In this case we will show that there is an extended region of the
eigenvalue density that contributes to the chiral condensate.

For simplicity, here we will show how the condensate arises from the
microscopic limit of the spectral density which is believed to 
be universal. We have also verified \cite{OSV2}
that a similar mechanism occurs even for larger $\mu$
provided that the contributing eigenvalues are still in the universal
region. 
The microscopic limit of the spectral density is defined as \cite{SV}
\be
&&\hat\rho_{N_f}(\hat x,\hat y,\hat m;\hat \mu) \\
&&= \lim_{V\to\infty}
\frac{1}{(\Sigma V)^2} \,\rho_{N_f}(\frac{\hat x}{\Sigma V},
\frac{\hat y}{\Sigma V},
\frac{\hat m}{\Sigma V}; \frac{\hat\mu}{F_\pi \sqrt{V}})\hspace*{0.5cm}\nn
\hspace*{-0.5cm}
\ee
In this limit,
$\hat x=x\Sigma V$, $\hat y=y\Sigma V$, $\hat m=m\Sigma V$ and 
$\hat\mu=\mu F_\pi \sqrt{V}$ are kept fixed with $\Sigma$ 
given by (\ref{psibarpsi}) and $F_\pi$ the pion 
decay constant.
The expression for the condensate (\ref{psibarpsi}) in this limit
becomes 
\be
\Sigma
= \lim_{\hat m,\hat\mu\to\infty}
\Sigma \int d{\hat x}\,d{\hat y} \,
\frac{\hat\rho_{N_f}(\hat x,\hat y,\hat m;\hat\mu)}{\hat x+i\hat y +\hat m}.
\label{psibarpsimicro}
\ee
The microscopic limit of the spectral density at nonzero
chemical potential 
was recently calculated both for the quenched case \cite{SplitVerb2} and the
unquenched case \cite {O}.
For a nonzero number of flavors it was found \cite{AOSV} 
that the eigenvalue density 
for $m\Sigma<2\mu^2F_\pi^2$
is a strongly oscillating complex function.
The oscillations cover a
region of the complex eigenvalue plane and, as we will see below,
the entire region 
contributes to the integral in (\ref{psibarpsimicro}). 
This constitutes a new mechanism where a discontinuity of the chiral
condensate in the complex mass plane is obtained from an oscillating
eigenvalue density in the complex plane. 
This mechanism does not rely on the specific form of the 
eigenvalue density as is demonstrated in the simple example below.

The lack of Hermiticity properties of the Dirac operator
at nonzero chemical potential is 
a direct consequence of the
imbalance between quarks and anti-quarks imposed
in order to induce a
nonzero baryon density. Because of the phase of the fermion
determinant, probabilistic methods are no longer effective
in the analysis of the partition function. This is known as
the sign problem. Although progress has been made in some
areas we believe that 
because of its physical origin, a paradigm shift will be necessary to develop
viable probabilistic algorithms for this problem.
Because of this, it is our opinion that it is particularly important
to improve our analytical understanding of chiral symmetry breaking
for QCD at nonzero baryon density.

Euclidean QCD at finite baryon density is not the only system 
without Hermiticity properties that has received much attention 
recently. 
We mention  the distribution of the poles of  $S-$matrices which are given by
the eigenvalues of a non-Hermitian operator \cite{mahaux,fyodorov},
the Hatano-Nelson model \cite{hatano-nelson}
(a random potential together with a nonzero  imaginary vector potential),
and
the description of  Laplacian growth 
in terms of the spectrum of non-Hermitian random matrices
\cite{Wiegmann}.  
The essential difference from QCD is that in these problems 
the determinant of the operator only enters in the generating function
of the resolvent. 
We will see that the additional determinant in QCD completely changes 
the character of the theory.

\vspace{2mm}

\noindent{\it Example.} As an example to illustrate our point, let us 
consider the eigenvalue density (${\rm sg}$ is the sign function)
\be
&&\hat \rho_{\rm Ex}(\hat x,\hat y,\hat m;\hat \mu)  =
 \frac {\theta(2\hat\mu^2-|\hat x|)}{4\pi\hat\mu^2 }
 \\&& \times
\left [1- e^{i {{\rm sg}(\hat x) (|\hat x| + 2\hat\mu^2) \hat y}/{4\hat\mu^2} 
+{(|\hat x|-|\hat m|)(|\hat x| + 2\hat\mu^2)}/{4\hat\mu^2}}\right ].\nn
\label{rhoex}
\ee
\begin{figure}[ht]
  \unitlength1.0cm
  \begin{center}
  \begin{picture}(3.0,2.0)
  \put(-2.,-4.5){
    \epsfig{file=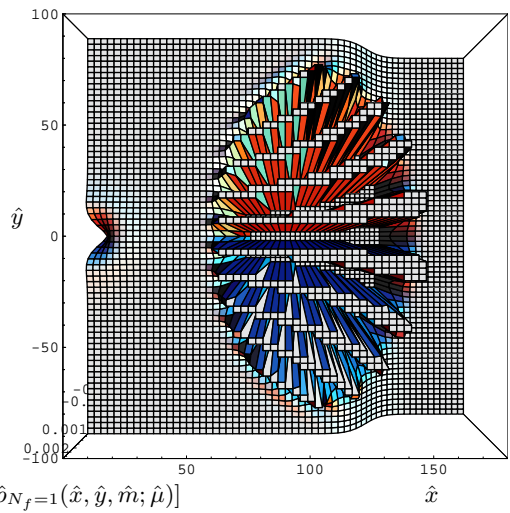,clip=,width=6.5cm}}
  \put(3.5,-4.7){\bf $\hat x$}
  \put(-2,-1.){\bf $\hat y$}
  \put(-2.7,-4.7){\bf ${\rm Re}[\hat\rho_{N_f=1}(\hat x,\hat y,\hat m;\hat\mu)]$}
  \end{picture}
  \vspace{4.5cm}
  \caption{ 
  \label{fig:ReRhoa1m100mu10}  
  Top view of the real part of the eigenvalue density for one flavor with
  $\hat m=60$ and $\hat\mu=8$ in half of the complex eigenvalue plane. The
  oscillations are cut off to illustrate the eigenvalue repulsion 
  at $\hat x=\hat y=0$ and the drop off for $\hat x>2\hat\mu^2$.}
  \vspace{-6mm}
  \end{center}
\end{figure}
This eigenvalue density has the same reflection symmetries as the
eigenvalue density of the QCD Dirac operator and 
has the property that it vanishes at the point where the
fermion determinant is zero.
 The integral in (\ref{psibarpsimicro}) can be evaluated analytically 
by means of a complex contour integral in $\hat y$ resulting in  
\be
\Sigma_{\rm Ex} = {\rm sg}(\hat m) \Sigma
+\frac \Sigma{\hat m}  e^{-|\hat m|} (e^{-|\hat m|}-1), 
\ee
which, for large $\hat m$, approaches ${\rm sg}(\hat m)\Sigma$. 
What we have learned from this example is that
a discontinuity in the chiral condensate  can be obtained 
from an oscillating 
spectral density rather
than from eigenvalues localized on the imaginary axis.
We will show next that the same mechanism is at work for QCD at
$\mu\ne 0$.

\vspace{2mm}

\noindent
{\it The microscopic spectral density.} The input for our 
calculation of the chiral condensate is the microscopic spectral density 
derived for any number of flavors in \cite{O}.
Without loss of generality we will 
consider the case $N_f=1$ and topological charge equal to zero for which 
the equations are less extensive. 
The microscopic eigenvalue density can be decomposed as \cite{O,AOSV}
\be
\hat\rho_{N_f=1}(\hat x,\hat y,\hat m;\hat\mu) =
 \hat\rho_Q(\hat x,\hat y;\hat\mu) - \hat\rho_U(\hat x,\hat y,\hat m;\hat\mu),
\label{dens-split}
\ee
with ($\hat z = \hat x+i\hat y$)
\be\label{rhoUontheway}
&&\hat\rho_U(\hat x,\hat y,\hat m;\hat\mu)= \frac{|\hat z|^2}{2\pi\hat\mu^2}
e^{-(\hat z^2+\hat z^{*\,2})/(8\hat\mu^2)}\\
&&\times 
K_0(\frac{|\hat z|^2}{4\hat\mu^2}) \frac{I_0(\hat z)}{I_0(\hat m)}
 \int_0^1 dt \, t e^{-2\hat\mu^2  t^2}
 I_0(\hat z^*t) I_0(\hat mt).\nn
\ee
The first term in (\ref{dens-split}) is the quenched eigenvalue 
density \cite{SplitVerb2} given by
$\hat\rho_Q(\hat x,\hat y;\hat\mu) =
 \hat\rho_U(\hat x,\hat y,\hat x+i\hat y;\hat\mu)$.
As expected, the microscopic spectral density vanishes at $\hat z=\pm \hat m$. 
A plot of the real part of the eigenvalue density for $\hat m=60$ 
and $\hat\mu=8$ is shown in figure \ref{fig:ReRhoa1m100mu10}.
Notice that the oscillatory region extends from the mass pole at 
$\hat z=\hat m$ and toward the boundary of the support of the spectrum.

The oscillations appear as the microscopic variables become
large, i.e.~as the thermodynamic limit is approached. In this region an
asymptotic formula for the eigenvalue density is accurate and will 
be used in order to analyze the role of the oscillations for chiral symmetry
breaking. We first derive the asymptotic formula for the
eigenvalue density and then evaluate the chiral condensate from
(\ref{psibarpsimicro}).   
For $\hat\mu^2\gg 1$ and $(\hat x+\hat m)/(4\hat\mu^2)<1$, the
integral in (\ref{rhoUontheway}) over $t$ is very well approximated by
\cite{gernot}
\be\label{approxbes}
&& \int_0^1
 dt\;  t\, e^{-2\hat\mu^2 t^2} I_0(\hat z^*t)I_0(\hat mt) \\ 
&&\approx \frac 1{4\hat\mu^2}\exp(\frac
{\hat z^{* \, 2}+\hat m^2}{8\hat\mu^2})   
I_0(\frac{\hat m \hat z^*}{4\hat\mu^2}).\nn
\ee
Furthermore, we are interested in the approach to the thermodynamic limit
where $|\hat m| \gg 1$, $|\hat z| \gg 1$, 
$|\hat z|^2/(4\hat\mu^2)\gg 1$ and $|\hat m \hat z|/(4\hat\mu^2)\gg 1$.
This justifies the replacement of the Bessel functions by their leading 
order asymptotic expansion including the Stokes terms.
We obtain the following asymptotic result for the
difference between the quenched and the unquenched eigenvalue density
\be\label{rhoUasymp}
&&\hat\rho_U(\hat x,\hat y,\hat m;\hat\mu) \\
&& \sim \frac{1}{4\pi\hat\mu^2 } 
e^{-[\hat y+i \,{\rm sg}(\hat x)(|\hat x|+|\hat m|-4\hat\mu^2)]^2/(8\hat\mu^2) 
-(|\hat x|-2\hat\mu^2)^2/(2\hat\mu^2)}.\nn
\label{rhouas}
\ee
This expression has the reflection symmetries discussed
below (\ref{rhoNf-def}). The asymptotic expansion of the quenched
part of the spectral density is simply given by
\be
\hat\rho_Q(\hat x,\hat y;\hat\mu)& \sim &
 \frac{1}{4\pi\hat\mu^2}\theta(2\hat\mu^2 - |\hat x|). 
\label{rhoqas}
\ee
For the argument presented below it is important that also 
the asymptotic expansion of the spectral density vanishes 
at $\hat x+i \hat y = \pm \hat m$.

\vspace{2mm}

\noindent
{\it The chiral condensate.} As explained in the introduction, 
the chiral condensate does not depend on the baryon chemical potential.
Hence, in the microscopic limit it is known that \cite{GLeps,LS} 
(momentarily we use the original variables to emphasize the volume
dependence) 
\be\label{Sigma-micro}
\Sigma_{N_f=1}(m) = 
\Sigma\frac{I_1(mV\Sigma)}{I_0(mV\Sigma)},
\ee
which is discontinuous, $\Sigma_{N_f=1}(m)={\rm sg}(m)\Sigma$, 
in the thermodynamic limit at fixed quark mass.
The question we wish to answer is how the oscillatory spectral
density conspires into a $\mu$ independent condensate. 
We stress that this is not just a challenging 
mathematical problem; understanding which parts of the eigenvalue density 
contributes to the chiral condensate will give direct insight in the 
physical consequences of the sign problem.

We now derive the chiral condensate from (\ref{psibarpsimicro}) using the 
asymptotic form of the microscopic eigenvalue density.
We first consider the integral over $\hat y$. The contribution 
from the quenched part of the spectral density 
(\ref{rhoqas}) is given by
\be\label{quenched-part}
\frac{1}{4\pi\hat\mu^2}\int_{-\infty}^\infty {\rm d}\hat y \
\frac{1}{\hat x+i\hat y+\hat m}=
 {\rm sg}(\hat x+\hat m)\frac{1}{4\hat \mu^2}.
\ee
The contribution from $\hat\rho_U$ in (\ref{psibarpsimicro}) is evaluated by 
a saddle point approximation. 
The contour in the complex $\hat y$ plane is deformed into a contour 
from $-\infty$ to $\infty$  over 
the saddle point at $\hat y =i\,{\rm sg}(\hat x)(4\hat\mu^2-|\hat x|-|\hat m|)$
and, if the contour has crossed the pole,  an integral around the
pole at $ \hat y = i(\hat x+\hat m)$. The saddle point contribution
is exponentially suppressed for $|\hat x| < 2\hat\mu^2$
leaving only the integral around the pole. For $\hat m > 0$ ($\hat m<0$)
the pole contribution for $\hat x >0$ ($\hat x<0$) is exponentially suppressed.
We obtain
\be\label{intpsibarpsi2}
&& \! \!\! \!\! \frac{1}{4\pi\hat\mu^2}\int_{-\infty}^\infty \!\! {\rm d}\hat y
e^{-[\hat y+i \,{\rm sg}(\hat x) (|\hat x|+|\hat m| -4\hat\mu^2)]^2/8\hat\mu^2 
-(|\hat x|-2\hat\mu^2)^2/2\hat\mu^2}\nn\\
&&\hspace{2cm} \times\frac{1}{\hat x+i\hat y+\hat m}  \\
&&\simeq  -\frac{1}{2\hat\mu^2}[\theta(\hat m)\theta(-\hat x-\hat m)
-\theta(-\hat m)\theta(\hat x+\hat m)],\nn
\ee
where we have used that the exponent vanishes at the pole.
For $\hat x>2\hat\mu^2$ the eigenvalue density is zero so it is now trivial
to do the integration over $\hat x$ to get 
\be
\Sigma_{N_f=1} &=& \frac{\Sigma}{2\hat\mu^2} \int_{-2\hat\mu^2}^{2\hat\mu^2}
{\rm d}\hat x \big[ \frac{1}{2}{\rm sg}(\hat x+\hat m) \\
&&  \ \ \ + \theta(\hat m)\theta(-\hat x-\hat m)
-\theta(-\hat m)\theta(\hat x+\hat m)\big]  \nn\\
& = & {\rm sg}(\hat m)\Sigma. \nn
\ee
This result agrees with (\ref{Sigma-micro}) for $|\hat m|\gg1$ where 
the asymptotic expansion of $\Sigma_{N_f=1}(m)$ is valid. Using
the exact microscopic spectral density we would have recovered the
mass dependence of (\ref{Sigma-micro}). 

We also emphasize that a finite result for the chiral condensate is
not obtained due to a cancellation of the pole and a zero of the 
fermion determinant. The pole term gives a finite contribution
for each of the two terms in (\ref{dens-split}) which do $not$ vanish at
$\hat z=\hat m$. We have checked numerically that the same mechanism results
in a discontinuity of the chiral condensate for more than one flavor.

The contribution from the unquenched part of the eigenvalue density to the
chiral condensate is dominated by the pole term because the 
exponential in
(\ref{intpsibarpsi2}) suppresses the integrand at the saddle
point in the complex $\hat y$-plane.
The simple result (\ref{intpsibarpsi2}) which
implies a nonzero chiral condensate in the chiral limit is thus directly
related to the complex phase of the spectral density.
The oscillating exponential has to suppress terms that diverge
exponentially with the volume which is achieved by
oscillations in the $\hat y$-direction 
with a period that scales as the inverse of the volume.
In contrast, for quenched QCD, 
where the eigenvalue density is 
real and positive, the chiral condensate vanishes 
in the chiral limit provided
that $\mu\neq 0$ as follows immediately from (\ref{quenched-part}). 
Instead a diquark condensate forms and breaks chiral symmetry just like in
phase quenched theories (see e.g. \cite{misha,dominique-JV}). Chiral 
symmetry breaking in phase quenched and full QCD occur by means of 
two different
mechanisms, and the oscillations of the eigenvalue density due to the sign
problem in the unquenched case distinguish between 
the two.   

\vspace{2mm}

\noindent
{\it The resolvent.} 
One can easily convince oneself that for $\hat z \ne \hat m$ the exponent in 
the integrand of the resolvent
\be
\hat G_{N_f}(\hat z,\hat z^*,\hat m;\hat \mu) = \Sigma \int d^2 \hat u
\frac{\hat\rho_{N_f}(\re(\hat u),\im(\hat u),\hat m;\hat\mu)}{\hat u + \hat z} 
~\label{resolvent}
\ee
results in expressions that diverge in the thermodynamic limit. Indeed, this 
was concluded from earlier Random Matrix calculations of the resolvent
\cite{HJV}. 
Using the microscopic spectral density (\ref{rhoUontheway}) one finds
good numerical agreement with the results obtained in \cite{HJV}.

\vspace{2mm}

\noindent
{\it Phase transitions in generating functions.} 
Our results can be understood in terms of phase transitions for
generating functions for the spectral density. Using the replica trick 
\cite{Girko,misha,AOSV}       
\be\label{replica}
\rho_{N_f}(x,y,m;\mu)
 =\lim_{n\to0} \frac{1}{\pi n}\partial_{z^*}\partial_z
 \log{\cal  Z}_{N_f,n}(m,z,z^*;\mu) \nn
\ee
we are naturally lead to the generating functionals 
\be
\label{repgenQCD}
{\cal Z}_{N_f,n}(m,z,z^*;\mu) = \langle  
\det(D+ m)^{N_f}|\det(D+z)|^{2n}\rangle \nn
\ee
for the eigenvalue density. The presence of conjugate quarks in the
generating function induces a coupling to the chemical
potential in the low-energy  effective theory which is completely
fixed by the pattern  of chiral symmetry breaking and determines
the eigenvalue density uniquely in the microscopic regime.   
In order to calculate the microscopic spectral density this way,
one has to employ powerful integrability relations that exist for the
effective partition function \cite{kanzieper02,SplitVerb1}.

Let us compare the phases of the generating function for  $n=1$ to
the regions in the complex plane characterized by a
different behavior of the eigenvalue density for $N_f =1$. 
For real $\hat z$ the phase structure follows from \cite{KT}. The 
extension for complex $\hat z$ can be obtained
from the asymptotics of the microscopic generating functions given 
in \cite{AOSV}.
For $|{\rm Re}(\hat z)|>2\hat\mu^2$ the generating functional is 
in the normal state while for $|{\rm Re}(\hat z)|<2\hat\mu^2$
two Bose condensates are separated by a first order phase transition.
This phase transition occurs exactly where the oscillating
region of the eigenvalue density begins, while the transition to the
normal phase corresponds to the drop off of the eigenvalue density for
$|{\rm Re}(\hat z)|>2\hat\mu^2$.

\vspace{2mm}

\noindent
{\it Conclusions.} 
For QCD at nonzero chemical potential
the chiral condensate is not dominated by the
contribution from the smallest eigenvalues.
On the contrary, we have found that the contributions from strips
parallel to the imaginary eigenvalue axis do not depend on 
the real part of the eigenvalue 
as long the
eigenvalue is inside  the support of the spectrum. This result 
arises from integrating a spectral density that oscillates with
a period of $1/(\Sigma V)$ and an amplitude that diverges exponentially
with the volume. 
Although we have shown only results using the microscopic eigenvalue density,
we have checked \cite{OSV2}
that our arguments apply up to $\mu \sim 1/L$ (for $V=L^4$). 
In conclusion, we have uncovered
a novel mechanism of chiral symmetry breaking at nonzero chemical potential
where an oscillatory spectral density results in 
a discontinuity of the chiral condensate in the complex mass plane.

\noindent
{\bf Acknowledgments:} We wish to thank Gernot Akemann,
Ove Splittorff and Andrew Jackson for discussions. 
J.C.O. was supported in part by U.S. DOE grant DE-FC02-01ER41180,
and J.J.M.V. was supported in part by U.S. DOE grant DE-FG-88ER40388.

\end{document}